# Isotope enrichment in neon clusters grown in helium nanodroplets

Lukas Tiefenthaler,[1] Siegfried Kollotzek,[1] Michael Gatchell,[1,2] Klavs Hansen,[3] Paul Scheier,[1,a] Olof Echt [1,4,a]

[1] Institut für Ionenphysik und Angewandte Physik, Universität Innsbruck, A-6020 Innsbruck, Austria
[2] Department of Physics, Stockholm University, 106 91 Stockholm, Sweden
[3] Center for Joint Quantum Studies and Department of Physics, Tianjin University, Tianjin 300072, China
[4] Department of Physics, University of New Hampshire, Durham NH 03824, USA

[a] Electronic email: olof.echt@unh.edu, paul.scheier@uibk.ac.at
Correspondence during submission process: Echt, Olof <Olof.Echt@unh.edu>

**Abstract**
Neon cluster ions Ne$_s^+$ grown in pre-ionized, mass-to-charge selected helium nanodroplets (HNDs) reveal a strong enrichment of the heavy isotope $^{22}$Ne that depends on cluster size $s$ and the experimental conditions. For small sizes the enrichment is much larger than previously reported for bare neon clusters grown in nozzle expansions and subsequently ionized. The enrichment is traced to the massive evaporation of neon atoms in a collision cell that is used to strip helium from the HNDs. We derive a relation between the enrichment of $^{22}$Ne in the cluster ion and its corresponding depletion factor $F$ in the vapor phase. The value thus found for $F$ is in excellent agreement with a theoretical expression that relates isotopic fractionation in two-phase equilibria of atomic gases to the Debye temperature. Furthermore, the difference in zero-point energies between the two isotopes computed from $F$ agrees reasonably well with theoretical studies of neon cluster ions that include nuclear quantum effects in the harmonic approximation. Another fitting parameter provides an estimate for the size $s_i$ of the precursor of the observed Ne$_s^+$. The value is in satisfactory agreement with the size estimated by modeling the growth of Ne$_s^+$, and with lower and upper limits deduced from other experimental data. On the other hand, neon clusters grown in neutral HNDs that are subsequently ionized by electron bombardment exhibit no statistically significant isotope enrichment at all. The finding suggests that the extent of ionization-induced dissociation of clusters embedded in HNDs is considerably smaller than for bare clusters.

# 1. Introduction

The primary motivation behind the first successful attempts to produce clusters of hydrogen and other permanent gases in supersonic expansions was the idea that the directed injection of large clusters of deuterium or tritium into a thermonuclear plasma would offer advantages over the injection of $^2H_2$ or $^3H_2$ molecules.[1] Very soon researchers at the Karlsruhe Nuclear Research Center discovered that expansion of hydrogen gas that contained 0.2 % $^2H_2$ resulted in clusters with a three times enriched deuterium content.[2] Shortly thereafter the group designed a process for the enrichment of $^{235}U$ in an expanding supersonic jet as an alternative to the gaseous diffusion and centrifuge processes.[3]

Few other studies of isotope enrichment in atomic clusters formed in nozzle beams have been published ever since. Kay and Castleman reported an enrichment of heavy water by about 30 % in small water clusters formed in a supersonic expansion of a 50:50 mixture of $H_2O$ and $D_2O$; the neutral clusters were ionized by electron bombardment.[4] Scheier and Märk have investigated neon cluster ions formed in a nozzle expansion of pure neon and subsequent electron ionization.[5] The researchers reported an increase of the abundance of $^{22}Ne$ from its natural value $p_{22,0} = 0.0925$[6] to $p_{22} \approx 0.124$ for clusters containing 4 to 22 atoms; the light isotope $^{20}Ne$ was depleted from its natural value $p_{20,0} = 0.9048$ to $p_{20} = 0.8745$. In other words, the experimental enrichment factor $\phi$, defined as[4,7]

$$\phi = (p_{22}/p_{20})/(p_{22,0}/p_{20,0}) \qquad (1)$$

was about 1.4. For the dimer and trimer ions the enrichment was lower.

Ding and coworkers ionized Ne clusters, formed in a supersonic expansion, by threshold electron photo ion coincidence; they reported $p_{22} \approx 0.13$ for $Ne_s^+$, $s > 5$.[8] For the dimer and trimer ion they found slightly smaller values which seemed to converge to $\approx 0.13$ when the photon energy was increased from about 20.7 to 21.4 eV.

Johnson and coworkers formed cationic neon clusters by association of Ne atoms onto seed $Ne^+$ ions in an ionized supersonic expansion.[7] The dimer and trimer ions were enriched by a factor $\approx 14$. For larger sizes the value of $\phi$ dropped rapidly to about 1.5 beyond $Ne_7^+$.

The isotope enrichment observed in the various experiments may have different origins. Authors who have studied isotope enrichment in clusters formed in supersonic beams and subsequently ionized by electron bombardment have argued that the enrichment occurs during cluster growth, either because of isotope dependent reaction rates, or because differences in the zero-point energy which lead to different dissociation rates.[4,5] On the other hand, the dependence of the enrichment factor on the energy of the ionizing photons prompted Ding and coworkers to conclude that the isotope effect was caused, at least in part, by fragmentation following ionization.[8]

A different method for the synthesis of clusters has emerged during the last two decades, namely by doping large helium nanodroplets (HNDs) in a pickup cell; the doped HNDs are subsequently ionized by electron collision. The approach has been used to produce clusters of neon,[9,10] water,[11,12] and numerous other weakly bound systems (for a recent review see [13]). To the best of our knowledge, the question of isotope enrichment has not been addressed in any of those reports. We will demonstrate that no statistically significant isotope enrichment occurs when $Ne_s^+$ ions are formed by this approach.

However, we observe a large enrichment if another, recently developed method is used to synthesize $Ne_s^+$. In this approach, neutral HNDs are formed in a nozzle expansion, ionized by electron collisions, and sent into an electrostatic deflector which transmits $He_N^{z+}$ ions with a specific size-to-charge ratio $N/z$.[14-16] The charged HNDs are then passed through a pickup cell filled with a dopant gas. In the final step the charged, doped HNDs are passed through a collision cell filled with He gas at ambient temperature in order to strip them of excess helium. In the process, a weakly bound dopant cluster ion will also evaporate atoms.

When analyzing mass spectra of neon cluster ions produced with this apparatus we noticed an enhancement of the heavy $^{22}Ne$ isotope that was much larger than previously observed with other approaches (except for the enrichment in dimer and trimer ions formed by ion–molecule exchange reactions at low temperature).[7] In this contribution we provide a systematic study of isotope enrichment in neon clusters that are grown either in neutral or charged HNDs. The enrichment observed for charged HNDs will be explained by the large number of evaporations from the neon cluster ions in the collision cell. Conversely, the absence of isotope enrichment upon doping neutral HNDs implies that the embedded Ne clusters suffer much less severe fragmentation upon ionization than bare clusters.

We derive a relation between $p_{22}$ and the size $s$ that involves only two fitting parameters. Both parameters have physical meaning and can be critically tested. One of them is the initial size $s_i$ of the neon cluster ion that is embedded in the HNDs that enter the collision cell. The value properly falls between the lower and upper limits that we estimate from a measured mass spectrum, and with a value obtained by modeling the events in the pickup and collision cells. The other fit parameter describes the depletion of $^{22}$Ne in the vapor phase which is conventionally measured in vapor pressure experiments; it provides a link between cluster beam experiments and measurements of bulk properties in thermodynamic equilibrium. Our result agrees very well with the prediction of a theory that involves the Debye temperature. Furthermore, the depletion factor provides an estimate for the difference in zero-point energies between cluster ions composed of $^{20}$Ne and $^{22}$Ne, respectively. The result agrees reasonably well with a theoretical study of nuclear quantum effects in Ne$_s^+$ by Calvo et al.[17]

## 2. Experiment

HNDs are formed by supersonic expansion of pre-cooled helium (Linde, purity 99.9999 %, stagnation pressure 20 bar) through a nozzle (5.7 μm diameter, temperature 9.7 K) into vacuum. The expanding beam is skimmed by a conical skimmer (Beam Dynamics, Inc.). The HNDs contain an estimated number of $N = 1 \times 10^6$ helium atoms.[18,19] Their size distribution will be broad but their speed distribution is narrow, with an average speed $v_0 \approx 238$ m/s.[19] Therefore their kinetic energy is proportional to their mass and, barring significant contamination, their size $N$.

Two different methods are used in the present work to produce Ne$_s^+$. In the first, conventional one, the neutral HNDs traverse a 5 cm long, differentially pumped pick-up cell filled with neon. The beam of doped helium droplets is collimated and crossed by an electron beam with a nominal energy of 50 eV. Cations are accelerated into the extraction region of a reflectron time-of-flight mass spectrometer (Tofwerk AG, model HTOF) with an effective mass resolution $m/\Delta m = 3000$ at $m =$ 262 u ($\Delta m$ = full-width-at-half-maximum, FWHM). Further experimental details have been provided elsewhere.[20]

The second, recently designed approach is illustrated in Fig. 1.[14] The neutral HNDs are electron ionized at 70 eV and an emission current of 630 μA. Multiple collisions with electrons may result in highly charged HNDs although the Rayleigh instability limits the charge that a HND of size $N$ can carry without undergoing spontaneous fission.[16] The He$_N^{z+}$ ions are accelerated into an electrostatic quadrupole bender. Ions that are transmitted have a specific mass-to-charge ratio $m/z$ because an electrostatic field selects ions by the ratio $K/z$ where $K$ is their kinetic energy. Upon acceleration they gain a kinetic energy $ez\Delta V$ where $e$ is the elementary charge. Hence their total kinetic energy is $K = aN + bz$ where $a$ and $b$ are constants, and ions with the same $N/z$ have the same $K/z$. In summary, the quadruple bender will select narrow slices from the broad size distribution of neutral HNDs, one slice for each charge state $z$.

Two sets of data were recorded, one with the quadrupole bender set to transmit droplets with $N/z$ = 46700, and another one with $N/z$ = 75000. As discussed in more detail in the Supplementary Material, the Rayleigh instability limits the charge state of the "small HNDs" to $z = 1$ while the maximum charge state of the "large HNDs" equals $z_{max} = 4$. In the main text we will focus on results obtained with the small, singly charged HNDs but some results obtained with large, multiply charged HNDs will be included as well.

Following the quadrupole bender, the size selected HNDs pass through a pickup cell (length $L =$ 5 cm) filled with neon gas at ambient temperature. The pressure, measured with a cold cathode ionization gauge (Pfeiffer model IKR 251), was 0.029 Pa (all pressures given here are corrected for the sensitivity of the ion gauge which is specified as 0.24 for neon and 0.17 for helium). Collisions between the charged HNDs and neon will lead to capture of Ne atoms, growth of $Ne_s^+$ in the droplet, and evaporation of helium atoms from the droplet.

The doped HNDs that exit the pickup cell pass through three regions in which they are guided by a radio frequency (RF) field. The components are taken from a commercial Quadrupole TOF (Q-TOF) mass spectrometer (Q-TOF Ultima from Micromass, Waters). The first one, a RF hexapole (length $L = 26$ cm), is filled with helium gas at ambient temperature. This cell will be referred to as collision cell. Collisions between the gas and the doped HNDs will lead to evaporation of the helium and, eventually, shrinkage of the bare neon cluster ions.

The following two ion guides, a quadrupole mass filter and a differentially pumped RF-hexapole collision cell, were not used in the present work. The ions were then extracted into a commercial time-of-flight mass spectrometer equipped with a double reflectron in W configuration and a microchannel plate (MCP) detector (Micromass Q-TOF Ultima mass spectrometer, Waters). The mass resolution was 8000 at 262 u.

Mass spectra were evaluated by means of a custom-designed software that corrects for experimental artifacts such as background signal levels, non-gaussian peak shapes and mass drift over time.[21] The routine takes into account the isotope pattern of all ions that might contribute to a specific mass peak by fitting a simulated spectrum with defined contributions from specific atoms to the measured spectrum in order to retrieve the abundance of ions with a specific stoichiometry.

## 3. Results

The top three panels in Fig. 2 display mass spectra of neon cluster ions produced by doping small charged, size-selected HNDs (average size $N = 46700$) with Ne in the pickup cell. The Rayleigh instability limits the charge state of these HNDs to $z = 1$. The doped HNDs then passed through the collision cell filled with helium gas at ambient temperature and varying collision pressures $P_{He}$.

In Fig. 2a the collision pressure was 0.0797 Pa, high enough to strip the droplets of all the helium. The ion yield reaches a maximum at $Ne_{55}^+$ and drops abruptly after $s = 56$. This local anomaly in the ion abundance agrees with previously reported mass spectra of neutral neon clusters formed in supersonic expansions of neon gas and subsequently ionized by electrons;[22] it correlates with closure of the second icosahedral shell.

Upon increasing the collision pressure, the size distribution of $Ne_s^+$ shifts to smaller sizes as seen in panels b and c. Sixteen other spectra were recorded. At the highest collision pressure, 0.443 Pa, the size distribution (not shown) is approximately exponential with an average size $s_{av} \approx 3.6$.

Fig. 2d presents a narrow region of the mass spectrum from panel c, revealing the isotopologues of $Ne_{13}^+$. Ne has three naturally occurring isotopes: $^{20}Ne$ (mass 19.99244 u, natural abundance $p_{20,0}$ = 0.9048), $^{21}Ne$ (20.99385, $p_{21,0}$ = 0.0027), and $^{22}Ne$ (21.99137 u, $p_{22,0}$ = 0.0925).[6] Prominent mass peaks appear at approximately even mass numbers; they are due to $^{20}Ne_{13}^+$, $^{20}Ne_{12}^{22}Ne^+$, $^{20}Ne_{11}^{22}Ne_2^+$, etc.. Peaks at approximately odd mass numbers contain a single $^{21}Ne$. Contributions from ions containing two or more $^{21}Ne$ atoms are negligible.

Fig. 2e displays the same mass region for a mass spectrum of neutral HNDs that were doped with Ne and subsequently ionized by electron collisions at 50 eV. In this approach, bare $He_N^+$ form a prominent ion series;[10] they are marked by asterisks. Mass peaks that appear at the expected positions of $Ne_{13}^+$ isotopologues are marked by dots. There are a few other unassigned mass peaks that are due to impurities. Compared to Fig. 2d, the distribution of even-numbered $Ne_{13}^+$ isotopologues is clearly shifted to smaller masses.

The data in Figs. 2d and 2e were analyzed using a custom-designed software[21] in order to extract the relative contributions of Ne isotopes to $Ne_{13}^+$. The results are $(p_{20}, p_{21}, p_{22})$ = (0.8412, 0.0039, 0.1549) for doping charged HNDs, and (0.9047, 0.0107, 0.0846) for doping neutral HNDs. In the latter case, ion peaks beyond 267 u were ignored because the fit indicated large contributions from background peaks. A comparison of experimental data and fit is presented in the form of histograms in Fig. 3. The theoretical distribution of isotopologues for natural neon is included in Fig. 3b. The agreement between the measurement (solid bars) and the fitted distribution (hashed bars) of isotopologues in Fig. 3a (small charged HNDs) is excellent; the statistical uncertainties (±0.0001) of the fitted value of $p_{22}$ are correspondingly small. The quality of the fit in Fig. 3b (neutral HNDs) is poor. The measured distribution is clearly broader than the trinomial distribution. The fitted value of $p_{22}$ is slightly *smaller* than the value for natural neon, but the difference is not deemed statistically significant. In short, for $Ne_{13}^+$ the abundance $p_{22}$ of $^{22}Ne$ is strongly enriched when small charged HNDs are doped, but there is no significant enrichment when neutral HNDs are doped and subsequently ionized.

The data in Fig. 4a present the size dependence of $p_{22}$ deduced from experiments on small charged HNDs measured for three different pressures in the He collision cell. The $p_{22}$ values decrease with increasing size $s$ and decreasing pressure $p_{He}$. The enrichment factor $\phi$, computed from eq.1,[23] is indicated along the ordinate on the right. The solid lines represent the results of fitting the data with eq. 8.

Data obtained from doping large charged HNDs are displayed in Fig. 4b with their error bars; the averages evaluated for each value of $s$ are indicated by the heavy solid line. Also shown in Fig. 4b are results of previous experiments on Ne cluster ions by Scheier et al., DeLuca et al., and Fieber et al..[5,7,8] The dashed line indicates the natural abundance $p_{22,0} = 0.0925$.

The size dependence of the abundance $p_{22}$ was extracted from a total of 11 mass spectra for neon cluster ions $Ne_s^+$ grown in small charged HNDs for various helium pressures $P_{He}$. Fig. 5 shows the dependence of $p_{22}$ on $P_{He}$ for $s$ = 5, 10, 15, and 20. In order to reduce statistical scatter, each data point represents the average including adjacent sizes. The value for $s$ = 5, e.g., includes data for $s$ = 4 and 6. Also shown is $p_{22}$ for $Ne_9^+$ cluster ions grown in large charged HNDs. This value was averaged over the size range $6 \leq s \leq 12$. By and large, the enrichment increases with increasing He pressure, decreasing size $s$, and it is largest when the cluster ions are grown in large HNDs.

## 4. Modeling Growth, Shrinkage and Isotope Enrichment in Neon Clusters

In the Discussion we will attribute the enrichment of $^{22}Ne$ in $Ne_s^+$ cluster ions formed in charged HNDs to the fact that the observed ions are the products of massive collision-induced dissociation of much larger parent clusters in the collision cell. In support of this scenario we will derive a relation between the observed enrichment of $^{22}Ne$ in the cluster ion and its corresponding depletion in the ensemble of evaporated atoms. The relation involves just two adjustable parameters. One of them is the initial size $s_i$ of the neon cluster ion that is embedded in the HNDs that enter the collision cell. This quantity cannot be measured directly; therefore we develop a model in Section 4.2 that allows to estimate its value. The model can be independently tested by modeling (in Section 4.3) the size of the neon cluster ion that will exit the collision cell and comparing it with its mass spectrometric value.

We will assume that the HNDs selected by the quadrupole bender are singly charged, i.e. only one singly charged $Ne_s^+$ will be embedded in each HND. In the experiment, multiply charged HNDs are excluded by selecting $He_N^{z+}$ with $N/z$ below $5.0 \times 10^4$ because the minimum size of doubly charged HNDs equals $1.00 \times 10^5$.[16] Our experiments with "small" charged HNDs meet this condition but experiments with "large" HNDs do not. In the Supplementary Material we will discuss the fate of "large" HNDs whose maximum charge state equals $z = 4$.

## 4.1 Enrichment of $^{22}Ne$ in the Cluster and its Depletion in the Vapor Phase

In this section we will derive a relation between the enrichment of $^{22}Ne$ in the cluster (which is measured in the present work) and its depletion in the ensemble of evaporated atoms. Let's consider just the $^{20}Ne$ and $^{22}Ne$ isotopes; the $^{21}Ne$ isotope (natural abundance 0.0027) is neglected. They are both bosons and therefore have the same quantum statistics, so we can forget any effect of this when considering relative values. The enrichment in the cluster is presumed to be by the slightly biased evaporation of one isotope over the other. To find the bias one can determine the relative rate constants $f_{20}$ and $f_{22}$ for the hypothetical case of isotopically pure clusters. There are three places the mass plays a role. One is in the zero-point energy of the nuclear motion. This will give the heavy isotope a higher dissociation energy and consequently a lower decay rate at identical temperatures.

As the branching ratios will be calculated below for evaporation from a single cluster, the assumption of identical temperatures is justified. The second effect is that the vibrational motion has a higher entropy for the heavy isotope. For each mode the reduction in quantum energy varies with the square root of the mass. Considering the clusters as harmonic oscillators, and assuming all modes are in the classical, high temperature limit, this gives a total reduction of the heavy to light isotope evaporation rate which is the mass ratio to the power 3/2, because an atom lost is three oscillators lost. The third factor is the phase space of the atoms, which enters the rate constant as another factor. It is proportional to the mass. The ratio of heavy to light evaporation rate constants is therefore the square root of the light to heavy mass ratio times the effective Boltzmann factor of the light to heavy isotope vibrational zero-point energy difference $\Delta E_{zp}$,

$$F = \frac{f_{22}}{f_{20}} = \sqrt{\frac{20}{22}} \exp\left(-\frac{D_{0,22}-D_{0,20}}{k_B T}\right) = \sqrt{\frac{20}{22}} \exp\left(-\frac{-E_{zp,22}+E_{zp,20}}{k_B T}\right) = \sqrt{\frac{20}{22}} \exp\left(-\frac{\Delta E_{zp}}{k_B T}\right) \quad (2)$$

Only the ratio of evaporation rates is of interest if one cares about the development with the number of lost atoms. Time does not enter this problem. If we ignore the variation of the Boltzmann factors with temperature, this ratio is a constant less than unity.

The ratio $f_{22}/f_{20}$ is not to be confused with the branching ratio $b_{22}/b_{20}$ for evaporation from an isotopically mixed cluster that contains $s_{20}$ and $s_{22}$ atoms of $^{20}$Ne and $^{22}$Ne, respectively, for a total of $s = s_{20} + s_{22}$. Nor should it be confused with the experimentally determined enrichment factor $\phi$ defined in eq. 1.[4] The $b$'s will be assumed to be proportional to the concentrations of the corresponding isotopes and the $f$'s. Hence the un-normalized $b$'s are

$$b_{2n} \propto f_{2n} \frac{s_{2n}}{s} \quad (3)$$

where $n$ is 2 or 0. Normalization to $b_{20} + b_{22} = 1$ gives

$$b_{2n} = \frac{f_{2n} s_{2n}}{f_{20} s_{20} + f_{22} s_{22}} \quad (4)$$

After one evaporation the concentration of the heavy isotope changes from its initial value $p_{22,0} = s_{22}/s$ to

$$p_{22} = \frac{s_{22}-b_{22}}{s-1} \approx \frac{s_{22}-b_{22}}{s}\left(1+\frac{1}{s}\right) \quad (5)$$

Keeping terms to leading and next-to-leading order in $1/s$ and equating discrete changes with derivatives allows us to write eq. 5 as

$$-\frac{d}{ds}\frac{s_{22}}{s} = -\frac{b_{22}}{s} + \frac{s_{22}}{s^2} = \frac{f_{22} s_{22}}{f_{20} s_{20} + f_{22} s_{22}} + \frac{s_{22}}{s^2} \quad (6)$$

To make life simple, the right-hand side will be approximated by setting $f_{22} = f_{20}$ in the denominator. This is not as poor an approximation as it appears at first sight because $f_{20} s_{20}$ remains the leading term. Then the differential equation can be written as

$$-\frac{s}{s_{22}}\frac{d}{ds}\frac{s_{22}}{s} = (1-F)\frac{1}{s} \quad (7)$$

Its solution

$$p_{22} = p_{22,0}\left(\frac{s_i}{s}\right)^{1-F} \quad (8)$$

provides the abundance of $^{22}$Ne in a cluster Ne$_s^+$ that was formed by multiple evaporations from Ne$_{si}^+$. A fit of this function to three data sets is shown in Fig. 4a. From the fit we obtain an average value $\langle s_i\rangle_{fit} = 124 \pm 5$. Further results will be discussed in Section 5; details are included in the Supplementary Material.

## 4.2 Growth of Neon Cluster Ions in the Pickup Cell

We will make the following assumptions:
4a. The HND is singly charged, with an initial size $N_i = 46700$.
4b. The sticking coefficient in the collisions between a neon atom and a charged HND equals 1.
4c. The capture cross section $\sigma$ of the droplet equals the hard-sphere value

$$\sigma = \pi(R_d + R_{Ne})^2 = \pi\left(R_{He} N^{1/3} + R_{Ne}\right)^2 = \pi R_{He}^2 \left(N^{1/3} + R_r^{-1}\right)^2 \quad (9)$$

where $R_d$ is the radius of the droplet which contains $N$ He atoms, $R_{He}$ and $R_{Ne}$ are the atomic radii of He and Ne, and $R_r = R_{He}/R_{Ne}$ characterizes the size of He relative to that of Ne. For helium we use $R_{He}$ = 0.222 nm;[24] for neon we calculate $R_{Ne}$ = 0.1744 nm from the molar volume, 13.39 cm$^3$/mol, of bulk neon at zero temperature and pressure.[25]

4d. The mass of the droplet which moves at speed $v_d$ through the scattering gas is much larger than that of a neon atom, i.e. the center of mass moves at speed $v_d$, and the reduced mass $m$ equals that of a neon atom.

4e. Each collision with Ne atoms releases an energy
$$E^* = E_t + D_{Ne} \qquad (10)$$
where $E_t$ is the energy transferred in the collision and $D_{Ne}$ the heat released upon addition of another Ne atom to the embedded neon cluster ion. $E_t$ equals $2 k_B T$ if the droplet is at rest. For a droplet moving through the thermal gas at speed $v_d$ one obtains[26]
$$E_t = 2k_B T + m_{Ne} v_d^2/2 \qquad (11)$$
$D_{Ne}$ equals the bulk heat of condensation, 20 meV. For our experimental conditions, each collision transfers $E^*$ = 104 meV to the droplet, except for the very early phase of Ne cluster growth to be discussed later. Here and elsewhere we neglect the effect of the weakly polarizable helium matrix on the energetics.

4f. Each collision causes the rapid evaporation of $E^*/D_{He}$ helium atoms where $D_{He}$ = 0.616 meV is the binding energy per atom in bulk helium,[27] and "rapid" means that the evaporation process is complete within a time much shorter than the transit time through the cell which is approximately 100 μs. Each collision results in the evaporation of 168 He atoms from the droplet. Non-thermal processes are neglected.

Assumption 4a is valid because the "small" HNDs selected by the quadrupole bender are smaller than the critical size of multiply charged HNDs.[16] Most of the other assumptions are rather obvious as long as the final number $N_f$ of helium atoms in the doped droplets that exit the pickup cell is much larger than the size $s$ of the embedded Ne$_s^+$. Eq. 9 will be modified when that assumption fails. Concerning assumption 4b we note that Lewerenz et al have measured the sticking coefficients in collisions of HNDs with Ar, Kr and Xe.[11] For Xe the value was only 0.58, but if one extrapolates their values to the mass of Ne one obtains approximately 1. Moreover, the droplets in their work contained only ≈2000 atoms, and they were neutral. We expect larger sticking coefficients if the droplets are larger, and charged.

We proceed to estimate the size of a neon cluster ion Ne$_s^+$ that grows in a HND that passes through the pickup cell. The collision frequency $f$ equals
$$f = n_{Ne} v_r \sigma \qquad (12)$$
where $n_{Ne}$ is the number density of Ne gas. The pressure was 0.029 Pa, hence $n_{Ne}$ = 6.9×10$^{18}$ m$^{-3}$. $v_r$ is the average relative collision speed in the center-of-mass system (which moves at speed $v_d$, see assumption 4d), i.e. |**v**-**v**$_d$| integrated over the surface of the droplet, all incident angles, and the thermal speed distribution of the scattering gas. The averaging can be simplified by noting that the asymptotic limit of $f$ equals $n_{Ne}\sigma v$ if the droplet moves slowly ($v_d \ll v$), and $n_{Ne}\sigma v_d$ if the droplet moves fast (because the droplet sweeps a volume $\sigma L$ within time $L/v_d$). We interpolate between the asymptotic limits with the expression
$$v_r \approx (v^2 + v_d^2)^{1/2} \qquad (13)$$
The number $dx$ of collisions with Ne in the pickup cell over a short path $dL$ may be written
$$dx = \frac{1}{\lambda} dL = \frac{f}{v_d} dL = n_{Ne}\sigma \frac{v_r}{v_d} dL \qquad (14)$$
where λ is the mean free path of the HND moving through the scattering gas.

Assumption 4f provides the relation between droplet shrinkage $dN$ and the number of collisions,
$$dN = -(E^*/D_{He}) dx \qquad (15)$$
which can be combined with eqs. 8 and 14 to write
$$\left(N^{1/3} + R_r^{-1}\right)^{-2} dN = -n_{Ne} \pi R_{He}^2 \frac{v_r}{v_d} \frac{E^*}{D_{He}} dL \qquad (16)$$

We have not yet taken into account the large energy release during the early stage of Ne cluster growth. Capture of the first Ne atom results in charge transfer between He$_3^+$, the ionic core in a charged HND,[28] and Ne. The reaction releases 0.50 eV, the difference between the adiabatic electron affinity of $He_3^+$ (22.07 eV)[29] and the ionization energy of Ne (21.566 eV).[8] Capture of a second Ne atom releases another 1.28 eV, the dissociation energy of $Ne_2^+$.[31] Capture of a third Ne atom releases another 0.10 eV, the dissociation energy of $Ne_3^+$.[32] Thus, formation of $Ne_3^+$ releases a total of 2.19 eV (this value includes the collisional contribution 3 $E_t$, see eq. 11), causing the loss of 3600 He atoms.

$Ne_3^+$ forms the ionic core in cationic neon clusters, hence Eq. 10 is reasonably accurate for capture of further Ne atoms.[33,34] By integration of eq. 16 using a reduced initial size $N_i' = N_i - 3600 = 43100$ and a reduced length $L' = L - 3\lambda = 4.8$ cm one finds $N_f = 29100$. From eq. 15 we conclude that the HNDs that exit the pickup cell contain an average of 86 neon atoms. Possible sources of error in this estimate will be considered at the end of the following Section.

### 4.3 The Fate of Doped HNDs in the Collision Cell

In this Section we model the fate of singly charged doped HNDs in the collision cell. The cell (length $L = 26$ cm) is filled with helium gas at $T = 300$ K and pressure $P_{He}$. The bulk evaporation energy of Ne is 32 times larger than that of helium, hence we can model the reactions in the collision cell in two phases: First, the ion will evaporate all its $N_i = 29100$ He atoms over a path length $L_1$; further collisions over the remaining path $L_2 = L - L_1$ will cause shrinkage of the bare Ne cluster ion which contains an initial number of about $<s_i>_{model} = 86$ atoms.

The model developed in Section 4.2 still applies, with the following changes: The radius $R_{Ne}$ in eq. 9, mass $m_{Ne}$ in eq. 11, and number density $n_{Ne}$ in eqs. 12, 14, 16 are to be replaced with the corresponding values for helium. The mass spectrum displayed in Fig. 2a which we will use to test our model was recorded with a helium pressure of $P_{He} = 0.0797$ Pa, hence $n_{He} = 1.92 \times 10^{19}$ m$^{-3}$. The thermal speed $v$ of helium and the relative collision speed $v_r$ (eq. 13) equal 1260 and 1375 m/s, respectively. Each collision releases an energy (cf. eqs. 10 and 11)

$$E^* = E_t = 2k_BT + m_{He}v_d^2/2 = 58.0 \text{ meV} \qquad (17)$$

This will lead to the evaporation of, on average, $E^*/D_{He} = 94$ He atoms per collision.

The expression for the collision cross section during phase 1 needs to take into account the contribution from the embedded $Ne_{86}^+$ which becomes non-negligible as the number of He atoms approaches zero. This is most easily done by noting that a Ne atom occupies the same volume as $(1/R_r)^3 = 0.488$ He atoms, and adjusting the integration limits accordingly. Thus the revised eq. 16

$$\left(N^{1/3} + 1\right)^{-2} dN = -n_{He}\pi R_{He}^2 \frac{v_r}{v_d} \frac{E^*}{D_{He}} dL \qquad (18)$$

is numerically integrated from an initial size $N_i' = N_i + 0.488 \ s = 29142$ to a final size $N_f' = 0 + 0.488 \ s = 42$. One finds that the ion will shed all its He atoms over a path length $L_1 = 10.1$ cm.

The final task is to determine the fate of bare $Ne_{86}^+$ during the remaining path $L_2 = L - L_1 = 15.9$ cm in the collision cell. Each collision will lead to an average loss of $E^*/D_{Ne} = 2.90$ Ne atoms. The collision cross section becomes

$$\sigma = \pi(R_d + R_{He})^2 = \pi(R_{Ne}s^{1/3} + R_{He})^2 = \pi R_{Ne}^2 (s^{1/3} + R_r)^2 \qquad (19)$$

and the revised eq. 16 becomes

$$\left(s^{1/3} + R_r\right)^{-2} ds = -n_{He}\pi R_{Ne}^2 \frac{v_r}{v_d} \frac{E^*}{D_{Ne}} dL \qquad (20)$$

which has the solution $<s_f>_{model} = 32$.

This value is significantly below the actual value $s_f = 66$ of the mass spectrum that we have attempted to model. The discrepancy suggests that $<s_i>_{model} = 86$ underestimates the size of the neon cluster ions that enter the collision cell. There are several potential sources of error to both our estimates, including the unknown accuracy of the pressure gauges and the correction factors used for measurements of helium and neon, collisions of the cluster ions with gas that leaks from the pickup cell and the collision cell into the vacuum chamber, and a gradual decrease of the drift speed $v_d$ due to the drag force. Furthermore, we have ignored the possible contribution of non-thermal processes to the shrinkage of the HND and, later, the neon clusters. Non-thermal processes are most likely when large amounts of energy are released, i.e. upon charge transfer from $He_2^+$ to Ne, and formation of $Ne_2^+$.

## 5. Discussion

We will start with a summary of results:

5a. For $Ne_s^+$ ions grown in charged HNDs, the abundance $p_{22}$ of $^{22}Ne$ is strongly enriched compared to its natural value $p_{0,22} = 0.0925$. For the neon dimer, $p_{22}$ ranges from 0.2 to 0.3. In contrast, no significant enrichment is observed when neutral HNDs are first doped and then ionized.

5b. For fixed experimental conditions, $p_{22}$ decreases with increasing size $s$.

5c. $p_{22}$ is largest when the charged HNDs are large.

The following results pertain to experiments with small charged HNDs that are singly charged. The pressure $P_{He}$ in the He collision cell was varied while the size of the HNDs and conditions in the Ne pickup cell remained unchanged.

5d. The value of $p_{22}$ increases with increasing pressure in the He collision cell.

5e. A reference mass spectrum recorded at low $P_{He}$ (0.0797 Pa) reveals a bell-shaped size distribution of $Ne_s^+$ with an average size $<s> = 66$ (see Fig. 2a). This value sets a lower limit to the size $s_i$ of Ne cluster ions (embedded in HNDs) that enter the He collision cell.

5f. We have modeled the fate of a charged HND in the pickup cell and the collision cell for the experimental conditions used to record the reference spectrum. According to the model, the initial and final sizes of the $Ne_s^+$ clusters that enter and exit the collision cell are $<s_i>_{model} = 86$ and $<s_f>_{model} = 32$, respectively.

5g. We have derived a relation between the cluster size dependence of $p_{22}$, the initial cluster size $s_i$, and the depletion factor $F = f_{22}/f_{20}$ where the $f$'s are the evaporation rate constants for the hypothetical case of isotopically pure clusters. A fit to all experimental data results in an average $<s_i>_{fit} = 124 \pm 5$. Values obtained for $F$ decrease from 0.82 for the smallest value of $P_{He}$ to about 0.75 for the largest pressure (see Supplementary Material).

5h. The fit factor $F$ depends on the difference in zero-point energies of the neon isotopes; values resulting from the fit can thus be tested.

We briefly review the pertinent literature. Scheier and Märk reported mass spectra of $Ne_s^+$, produced in a supersonic expansion of neon gas with subsequent electron ionization.[5] Their data are included in Fig. 4b (open squares). The value of $p_{22}$ increases from $s = 2$ to 5 and remains constant at $p_{22} \approx 0.124$ up to $s = 22$.

Also included in Fig. 4b are data reported by Ding and coworkers.[8] The authors measured $p_{22} \approx 0.13$ for $Ne_s^+$ ($s > 5$), formed in a supersonic expansion and ionized using a threshold-electron-photoion coincidence technique. For the dimer and trimer ion they found slightly smaller values but the values converged to ≈0.13 when the photon energy was increased from about 20.7 to 21.4 eV.

Kay and Castleman produced neutral water clusters in a supersonic expansion of a 50:50 mixture of $H_2O$ and $D_2O$ vapor; the neutral clusters were then ionized by electron bombardment.[4] They observed enrichment of $D_2O$ by about 30 to 40 %, about the same value as found for neon clusters. Note that the mass ratio of the heavy isotope and the light isotope is approximately the same for water and neon. In a study of krypton clusters formed in a supersonic expansion, for which the isotopic mass ratio is much smaller, Scheier and Märk did not find any isotope enrichment.[35]

Johnson and coworkers formed neon cluster ions by association of neutrals onto $Ne^+$ seed ions in an ionized supersonic expansion.[7] The dimer and trimer featured very large enrichment factors which were attributed to the low temperature achieved in the supersonic jet. Wei et al. found that the metastable fractions for monomer evaporation from fully deuterated ammonia cluster ions were higher than for non-deuterated ammonia cluster ions.[36] Fedor et al. measured the kinetic energy release distribution (KERD) upon metastable dissociation of $^{20}Ne_2^+$ and $^{22}Ne_2^+$.[37] They observed large differences in the KERD which they assigned to a mass-dependence in the relative rates of two competing electronic transitions. The reports mentioned in this paragraph have no direct relevance to our results, they will not be considered further.

Isotope enrichment in cluster beam experiments may have several possible origins. The two most likely ones are (1) enrichment during cluster growth, or (2) enrichment caused by dissociation. Less likely mechanisms are isotope-dependent ionization cross sections or detection efficiencies. Scheier and Märk have favored mechanism (1) because they did not observe any isotope dependence in the unimolecular dissociation rates.[5] The enrichment of $D_2O$ in water clusters was also attributed to mechanism (1) by Kay and Castleman.[4] On the other hand, Fieber et al. observed that $p_{22}$ increases when the energy of the ionizing photons was increased; they concluded that the enrichment of $^{22}$Ne is caused, at least in part, by ionization-induced dissociation.[8]

Although the mechanism for isotope enrichment may depend on the experimental approach, it is tempting to attribute its origin in the current experiments and previous reports[4,5,8] to a common cause, namely evaporation of atoms from $Ne_s^+$. In our approach we can safely exclude isotope enrichment during growth. Every atom that collides with the droplet is captured. The velocity of Ne atoms in the pickup cell greatly exceeds the critical Landau velocity of about 50 m/s below which the projectiles cannot create elementary excitations.[38]

Evaporation of $^{22}$Ne will be disfavored relative to that of $^{20}$Ne because the lower zero-point energy of $^{22}$Ne results in a larger quantum mechanical evaporation energy $D_{0,22}$, see eq. 2. The effective (vibrational) temperature of cluster ions in the gas phase depends primarily on their evaporation energies provided they were "boiling hot" when formed.[39] Hence, for a given system (neon), the value of $F$ will not depend on the experimental approach. Different values of $p_{22}$ in $Ne_s^+$ must then originate from differences in their precursor sizes $s_i$ or, more exactly, the ratio of the initial cluster size $s_i$ and the size $s$ of the observed cluster, see eq. 8.

The scenario readily explains most of the observations listed above. First, the $p_{22}$ values obtained here for neon clusters embedded in neutral and charged HNDs bracket those reported for ionization of bare clusters (observation 5a).[5,8] It is commonly agreed that ionization-induced dissociation of clusters embedded in neutral HNDs is less severe than for bare clusters, although details depend on nature of the dissociation channel.[40] In early reports it was even concluded that ionization-induced fragmentation can be completely quenched if the clusters are embedded in HNDs.[11] That assessment was too optimistic; cluster spectra feature the usual local abundance anomalies (magic numbers) that signal local anomalies in the stability of charged clusters of, e.g., neon (see Fig. 2a), argon, or krypton.[10,20,41] As the size distributions of their neutral precursors grown in HNDs are smooth, this implies that the cluster ions have lost at least one monomer.[42] After all, ionization of a neutral cluster always leads to the sudden release of a significant amount of excess energy upon formation of a tightly bound dimer or trimer ion in the neon cluster.[32,34,43] On the other hand, cluster ions grown in charged HNDs are forced to evaporate many atoms in the evaporation cell, leading to large $p_{22}$ values.

Second, $p_{22}$ values decrease with increasing size $s$ because the ratio $s_i/s$ decreases (thus explaining observation 5b). Third, doping large charged HNDs results in larger $p_{22}$ values because $s_i$ is larger (observation 5c). Fourth, increasing $P_{He}$ results in larger $p_{22}$ values because the initial size distribution of neon cluster ions is not strictly mono-disperse (observation 5d). For a given product size $s$, the value of the (average) precursor size $s_i$ increases as $P_{He}$ is increased.

We turn to a more quantitative discussion of our results. We start with the size $s_i$ of the cluster ions that enter the collision cell. It affects, according to eq. 8, the value of $p_{22}$. We cannot measure $s_i$ because without stripping helium from the doped, charged HND the ion mass would be too large. However, the mass spectrum in Fig. 2a establishes a lower limit, $s_{min} = 66$ (observation 5e). According to the model developed in Section 4.2, the doped HNDs that enter the collision cell contain $<s_i>_{model} = 86$ Ne atoms (observation 5f). On the other hand, a fit of eq. 8 to a total of 11 data sets of $p_{22}$ versus $s$, measured for 11 different He pressures, results in $<s_i>_{fit} = 124 \pm 5$ (observation 5g). There is no obvious dependence of the fitted values on the He pressure (which is, admittedly, at variance with our observation in the previous paragraph).

The width of the size distribution of the mass spectrum (Fig. 2a) provides another, although rough, way to estimate the value of $s_i$. The size distribution of $Ne_s^+$ grown in monodisperse charged HNDs follows a Poisson distribution whose width (standard deviation) equals $\sigma_{si} = \sqrt{s_i}$. Subsequent loss of $\Delta s = s_i - s$ atoms is another statistical process that broadens the distribution further by $\sigma_{\Delta s} = \sqrt{\Delta s}$; the expected width of the observed size distribution will be $\sigma_s = (\sigma_{si}^2 + \sigma_{\Delta s}^2)^{1/2} = (s_i + \Delta s)^{1/2} = (2s_i - s)^{1/2}$. This is a lower limit for the observed width because the size of the HNDs does not remain constant in the pickup cell, and loss of He in the collision cell may add to the broadening.

Turned around, the relation provides an upper limit for $s_i$. The mass spectrum in Fig. 2a is characterized by $s = 66$ and $\sigma_s = 13.2$. It follows that $s_i \leq s_{max} = 120$. The four estimates for $s_i$ or its limits span quite a range, from 66 to 124; the true value is probably somewhere between about 90 and 120, much larger than the size $s$ for which we have analyzed the value of $p_{22}$ (see Fig. 4a). The observed ions are the result of massive loss of Ne from the precursor ions.

The enrichment is even larger for $Ne_s^+$ cluster ions grown in larger HNDs, $N/z = 75000$. In these experiments, the Ne pressure in the pickup cell was also increased. A fit of eq. 8 to the average of those data (represented by the heavy line in Fig. 4b) results in $s_i \approx 400$. Our model developed in Sections 4.2 & 4.3 suggests $s_i \approx 300$. A detailed analysis is hampered by the fact that we are dealing with a superposition of precursor sizes because the charge state of HNDs with $N/z = 75000$ may be as large as $z = 4$. It is possible to model the fate of the cluster ions for each charge state. However, without knowing their relative contributions to the ensemble of $N/z$ selected HNDs we can merely state that these data confirm the trend: larger precursors result in larger enrichment.

We now turn to the other fit parameter, the depletion factor $F = f_{22}/f_{20}$ where the $f$'s are the evaporation rate constants for the hypothetical case of isotopically pure clusters. A fit of eq. 8 to eleven data sets recorded with small, singly charged HNDs shows that $F$ decreases from 0.82 for the smallest He pressure to about 0.75 for the largest pressure (see Supplementary Material). The uncertainty of the fit parameters grows as the He pressure grows because the range of sizes $s$ that can be fitted shrinks. $F \approx 0.80$ is a representative result for medium He pressures. From eq. 2, this value implies

$$\Delta E_{zp} = ln\left(\frac{1}{F}\sqrt{\frac{20}{22}}\right) k_B T = 0.176\, k_B T \tag{21}$$

The estimated vibrational temperature of medium-sized clusters (or cluster ions) subject to unimolecular dissociation equals $T \approx D/(\gamma\, k_B)$ where $D$ is the evaporation energy, and $\gamma \approx 25$ is the Gspann factor.[39] With $D \approx 20$ meV we obtain $T \approx 10$ K for neon. In our experiments, the evaporation rate is raised from $k \approx 10^4$ s$^{-1}$ in a typical metastable time window to $\approx 10^6$ s$^{-1}$. This decreases the Gspann factor by about 30 %,[44] hence we estimate $T \approx 13$ K. From eq. 21 we conclude that the difference in the zero-point energies of hypothetically pure $^{20}Ne_s^+$ and $^{22}Ne_s^+$ equals $\Delta E_{zp} \approx 0.20$ meV.

For comparison, we can estimate $\Delta E_{zp}$ for clusters from a theoretical study by Calvo et al..[17] The researchers determined the stable structures of $Ne_s^+$, $s \leq 57$ using a diatomic-in-molecules potential energy surface; the effects of vibrational delocalization were included in the harmonic approximation. For $s \geq 20$ the quantum mechanical dissociation energies were, on average, 6.7 meV smaller than the classical ones.[45] The zero-point energy scales as $\sqrt{1/m}$ where $m$ is the mass. Hence the zero-point energy of $^{22}Ne_s^+$ would be lower than that of $^{20}Ne_s^+$ by $\Delta E_{zp} = E_{zp}[1-\sqrt{(20/22)}] = 0.31$ meV, in reasonable agreement (observation 5h) with our value of 0.20 meV.

We can also compare our value $F = 0.80$ with direct measurements of isotope fractionation in the vapor phase of bulk neon.[46] The relation between fractionation and temperature is derived from the Clausius Clapeyron equation; it involves an integral of the difference in the heat capacities of the isotopes over the temperature.[47] Adopting the Debye theory of the heat capacity, the expression simplifies for atomic gases to

$$-\ln F = \frac{3}{40}\left(\frac{\theta}{T}\right)^2 \frac{m_{22}-m_{20}}{m_{22}} \tag{22}$$

The Debye temperature $\theta$ of neon in the low-temperature limit equals 74.6 K.[48] At 13 K eq. 22 predicts a depletion factor $F = 0.80$ in excellent agreement with the value obtained by fitting relation 8 to our experimental data.

One feature in our results that is left unexplained is the decrease in the value of $F$ with increasing He pressure. At the lowest pressure we find $F \approx 0.82$ which translates to $\Delta E_{zp} = 0.17$ meV; at highest pressure we find $F \approx 0.75$ which translates to $\Delta E_{zp} = 0.27$ meV. One possible factor is the shift in the average product size $s$ to smaller values as the He pressure increases. Small cluster ions are more strongly bound and their zero-point energies will be higher (for the dimer ion we estimate $\Delta E_{zp} = 1.7$ meV from density functional calculations by Michels et al.[49]). Another possible factor is that the fitting function eq. 8 does not allow for a size dependence of $F$. This is not consistent with eq. 2 because the cluster dissociation energy and hence the cluster temperature will increase with decreasing size $s$.

**Conclusion**
We have synthesized cationic Ne clusters by two different approaches. In the first, conventional one, neutral HNDs were doped with Ne and subsequently ionized by electrons. No significant enrichment was found for the heavy $^{22}$Ne isotope. In the second, novel approach, charged HNDs were doped with neon; excess helium was then removed by collisions of the doped HND ions at thermal energies in a collision cell filled with helium gas. Further collisions result in shrinkage of the $Ne_s^+$ cluster ions. The enrichment factor of $^{22}$Ne in the clusters was found to depend on the size $s$, the size of the precursor clusters, and the pressure in the He collision cell. The factor reached a value of $\approx 4$ in experiments with large HNDs.

Furthermore, in the second approach the charged HNDs were size-to-charge ($N/z$) selected. For small values of $N/z$ (which guarantee charge state $z = 1$), we have modeled the events in the pickup cell and collision cell and concluded that the dopant clusters in the HNDs that enter the collision cell contain about 86 Ne atoms. A somewhat larger value, $s_i \approx 124$, was deduced by fitting an analytical expression to the size dependence of the abundance of $^{22}$Ne. The only other parameter in the fitting function measures the depletion factor $F$ of $^{22}$Ne in the vapor phase. Its value has been checked against the fractionation determined in standard vapor pressure measurements,[47] and the calculated difference in the zero-point energies between cationic clusters of $^{20}$Ne and $^{22}$Ne.[17] In both cases, good agreement was obtained.

The growth of clusters in charged HNDs is a novel method with distinct features. The large enrichment of the heavy neon isotope reported here can be traced to the extensive evaporation cascade caused by numerous but gentle collisions with helium atoms in the collision cell. Even though roughly $10^2$ neon atoms are evaporated from the initial $Ne_s^+$, the temperature of the cluster ion remains at a constant value estimated to be 13 K. The temperature would be higher in experiments with more strongly bound clusters, but the large excess energies and athermal processes that typically accompany post-ionization of clusters, either bare or embedded in HNDs, are avoided.

**Supplementary Material**
See supplementary material for additional results, and for a discussion of the fate of multiply charged HNDs in the pickup and collision cells.

**Data Availability**
The data that support the findings of this study are available from the corresponding authors upon reasonable request.

Acknowledgement
This work was supported by the Austrian Science Fund, FWF (Projects I4130 and P31149) and the Swedish Research Council (Contract No. 2016-06625).

**Figures**

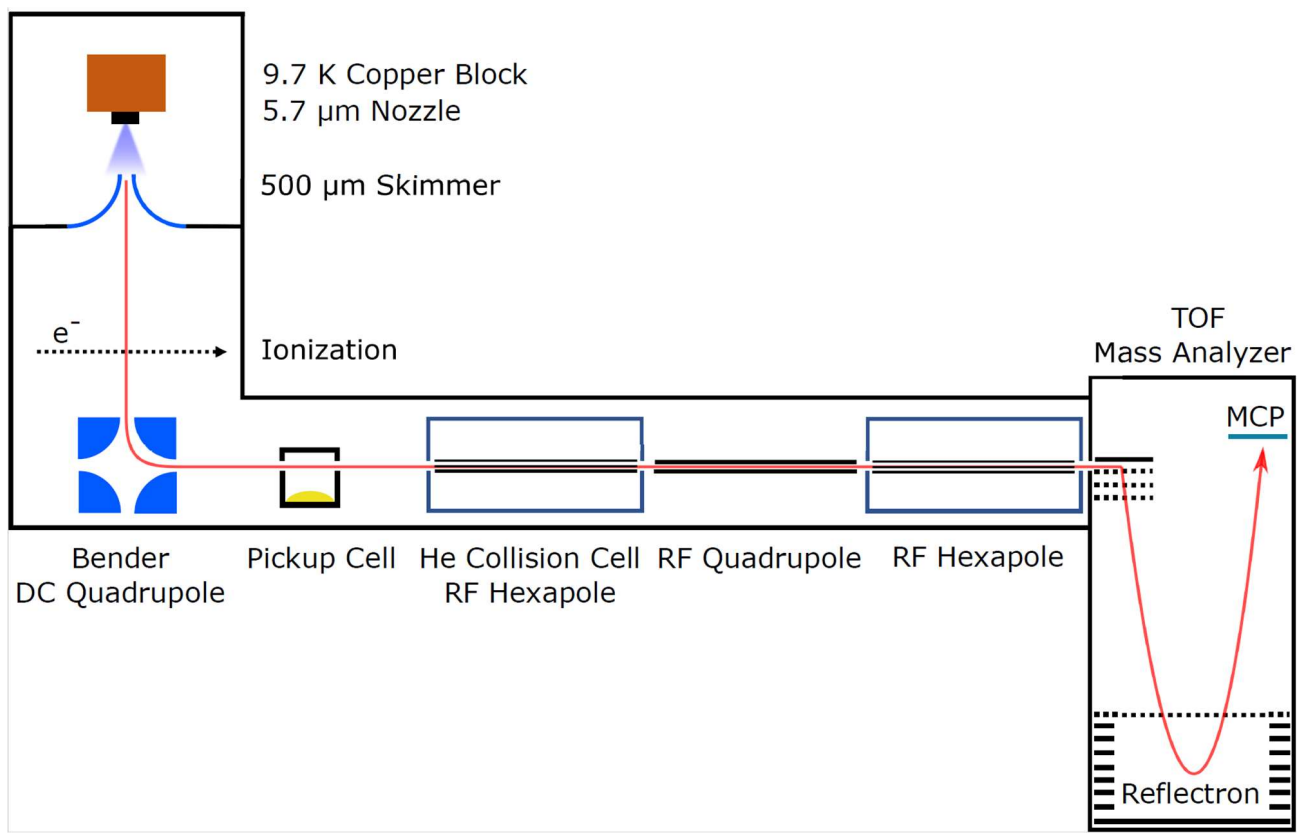

Fig. 1. The experimental setup used to dope size-to-charge selected cationic helium nanodroplets (HNDs) with neon. Bare Ne$_s^+$ cluster ions are then produced by collisions with He gas at ambient temperature in the collision cell.

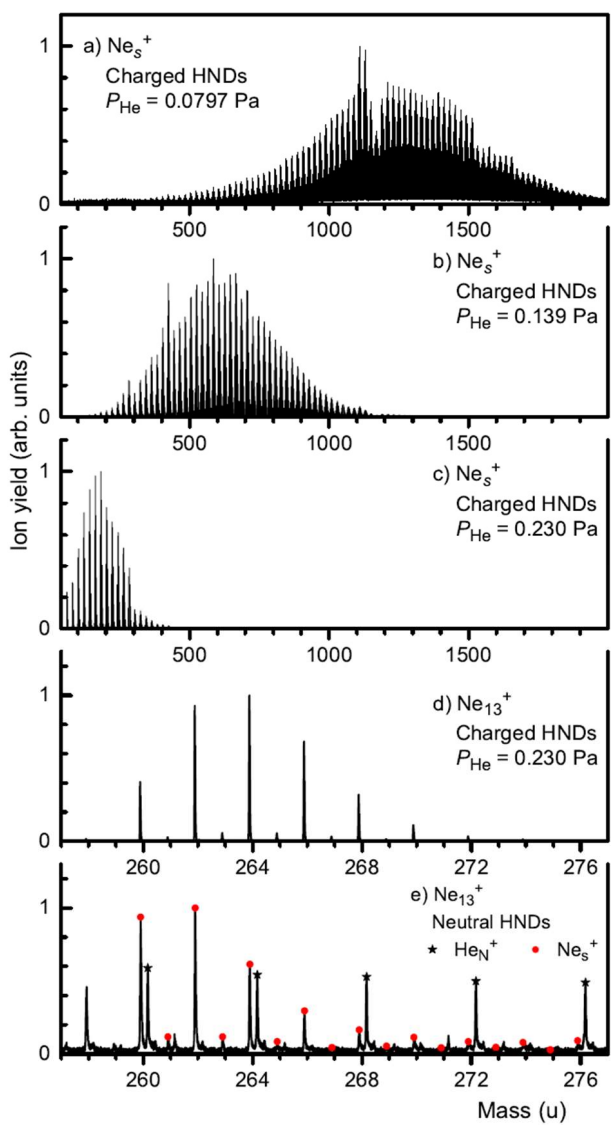

Fig. 2. Mass spectra of neon cluster ions $Ne_s^+$ grown in singly charged HNDs (panels a through d), and in neutral HNDs that were subsequently ionized by electrons (panel e). The spectra in panels a through c demonstrate the effect of increasing the helium pressure in the collision cell, leading to shrinkage of $Ne_s^+$. Panel d zooms into the spectrum shown in panel c in order to reveal the distribution of $Ne_{13}^+$ isotopologues. Panel e zooms into a spectrum of neon clusters grown in neutral HNDs and subsequently ionized. Peaks at the expected positions of $He_N^+$ and $Ne_{13}^+$ isotopologues are marked.

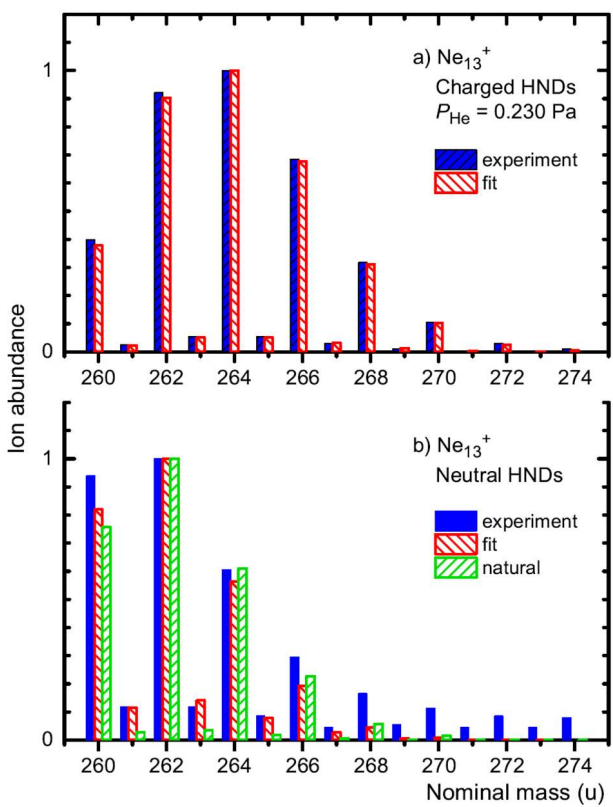

Fig. 3. Comparison of the measured yield of $Ne_{13}^+$ isotopologues deduced from the spectra displayed in Figs. 2d and 2e (full bars) with a fitted distribution of isotopologues (hashed bars). The isotope abundances $p_{21}$ and $p_{22}$ are fit parameters. Also shown in Fig. 3b is the distribution of isotopologues expected for a natural abundance of isotopes.

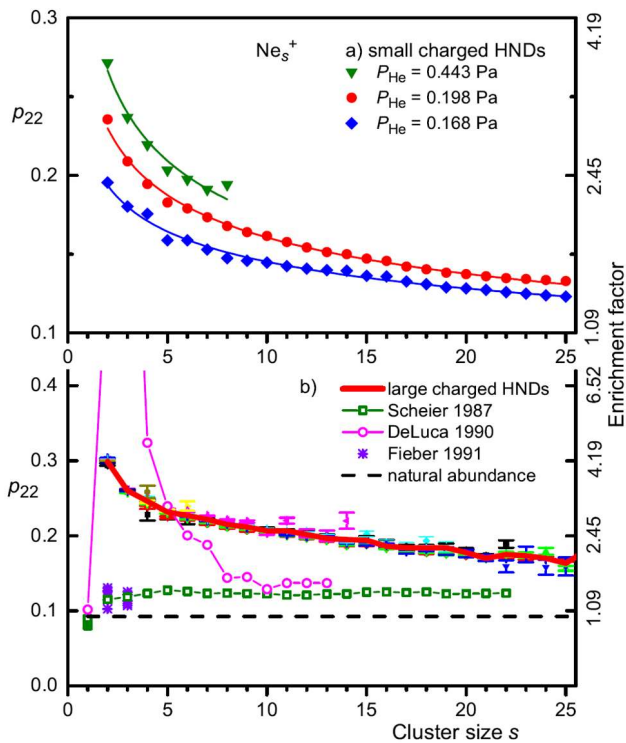

Fig. 4. Panel a: The size dependence of the isotope abundance $p_{22}$ in cluster ions $Ne_s^+$ that were grown in small, singly charged HNDs for three different pressures in the He collision cell. The solid lines result from a fit of eq. 8 to the data. Similar data obtained for $Ne_s^+$ grown in large, charged HNDs are displayed by data points with error bars in panel b; the heavy solid line represents the average. Also shown are the natural abundance of $^{22}$Ne (dashed line), values reported in the literature for ionization of bare clusters,[5,8] and for clusters grown by association of Ne atoms onto seed $Ne^+$ ions in an ionized supersonic expansion.[7] The enrichment factor $\phi$, computed from eq. 1, is indicated along the ordinate on the right.

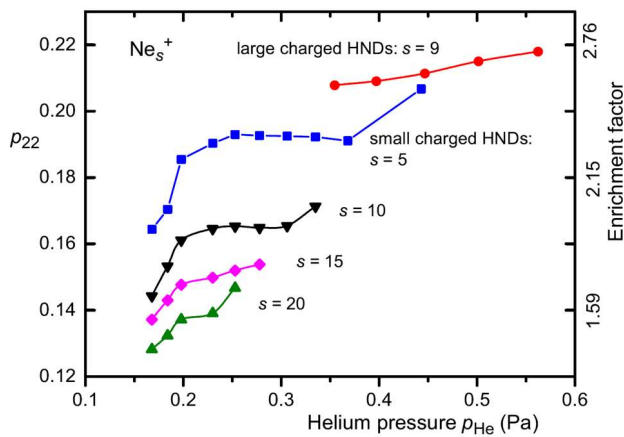

Fig. 5. The isotope abundance of $^{22}$Ne in $Ne_s^+$ ions grown in small, singly charged HNDs for four different sizes $s$ versus the helium pressure in the collision cell. Lines are drawn to guide the eye. Also shown is $p_{22}$ of $Ne_9^+$ ions grown in large charged HNDs. All data were averaged over a narrow range of $s$ values in order to reduce statistical scatter.

# Isotope enrichment in neon clusters grown in helium nanodroplets

## Supplementary Material

Lukas Tiefenthaler, Siegfried Kollotzek, Michael Gatchell, Klavs Hansen, Paul Scheier, Olof Echt

## S1   Additional Results

The abundance $p_{22}$ of $^{22}$Ne in the cluster ion is related to the depletion factor $F = f_{22}/f_{20}$ in the gas phase by the relation

$$p_{22} = p_{22,0} \left(\frac{s_i}{s}\right)^{1-F} \tag{S1}$$

where $s$ is the cluster size for which $p_{22}$ is measured, and $s_i$ the (average) size of its precursor.

Eq. S1 has been fitted to 11 data sets that were recorded for small, singly charged HNDs ($N/z = 46700$). The He pressures ranged from 0.168 to 0.443 Pa. For smaller pressures the neon clusters were too large for an analysis.

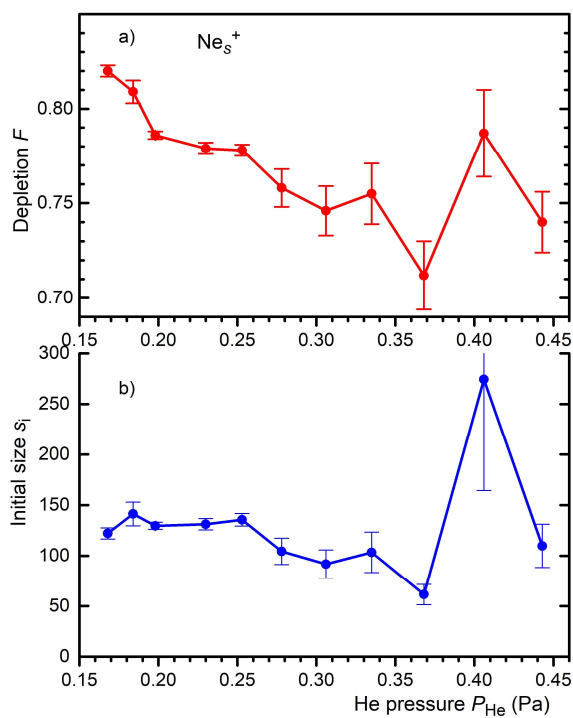

Fig. S1: Results from fitting eq. S1 to 11 data sets.

## S2 Multiply charged HNDs

Multiply charged $He_N^{z+}$ droplets become unstable with respect to fission when their size $N$ falls below the critical size $N_z = 3.54 \times 10^4 \, z^{3/2}$, i.e. $N_2 = 10^5$, $N_3 = 1.84 \times 10^5$, $2.83 \times 10^5$, etc.[1] Viewed another way, the Rayleigh instability sets an upper limit $z_{max}$ to the charge that a HND of size $N$ can support without undergoing spontaneous fission. For the "small" HNDs discussed in the main text, $N/z = 46700$ and $z_{max} = 1$. For the large HNDs, $N/z = 75000$ and $z_{max} = 4$. The Ne pressure in the pickup cell was 0.082 Pa, the He pressure in the collision cell was 0.112 Pa. Upon doping and collisions with helium gas the droplets will shrink, undergo fission, and eventually end up as singly charged $Ne_s^+$.

Experiments with undoped HNDs have shown that a very small singly charged helium cluster ion will be ejected upon fission, containing fewer than $\approx 10^2$ atoms.[1,2] These results pertain to undoped HNDs. They should equally well apply to doped clusters where the charges are localized on small $Ne_s^+$ rather than $He_2^+$ or $He_3^+$, provided the dopant cluster ions occupy a relatively small volume in the HND.

Two mass spectra recorded for large HNDs are shown in Fig. S2. Fig. S3 compares the measured distribution of isotopologues with fitted distributions.

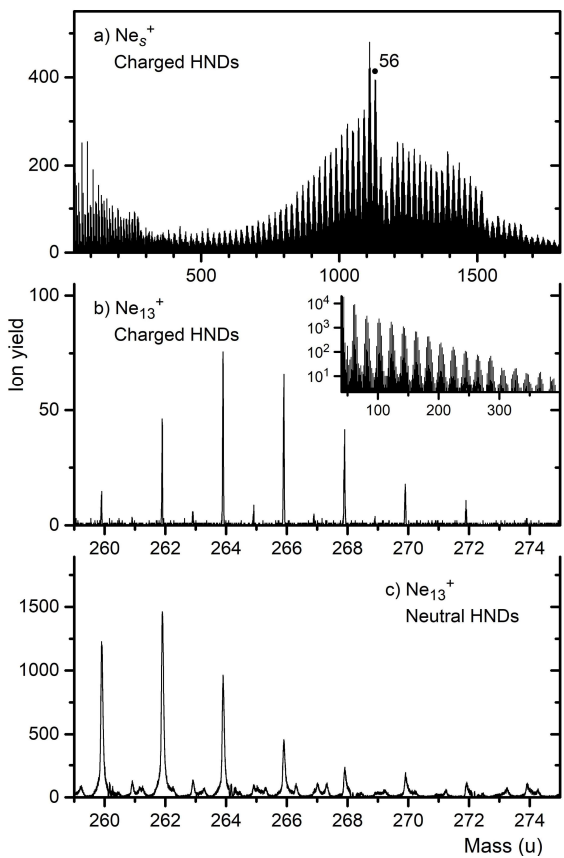

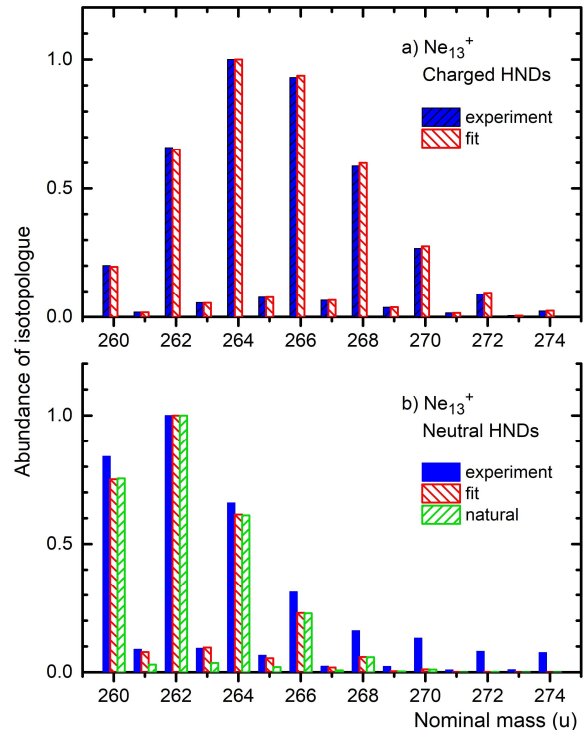

**Fig. S2**, Panel a: Mass spectrum of neon cluster ions $Ne_s^+$ grown in large charged HNDs. The inset in panel b shows a spectrum recorded at higher pressure in the He collision cell. Sections of mass spectra revealing the distribution of $Ne_{13}^+$ isotopologues for clusters grown in charged and neutral HNDs are displayed in panels b and c, respectively. A prominent series of pure $He_N^+$ that appears in spectra of neutral HNDs has been fitted and subtracted from the data in panel c.

**Fig. S3:** Comparison of the measured yield of $Ne_{13}^+$ isotopologues (blue bars, deduced from the spectra displayed in Figs. S2b and S3c) with a fitted distribution of isotopologues (hashed bars). The largest bar was arbitrarily set to 1; the isotope abundances $p_{21}$ and $p_{22}$ are fit parameters. Also shown in Fig. S3b is the distribution expected for a natural abundance of isotopes.

**S3 The Fate of Multiply Charged HNDs in the Pickup and Collision Cells**

The analysis in the main text assumed that the undoped HNDs are singly charged. Electron ionization of HNDs at high emission currents results in multiply charged HNDs with charge states as high as $z = 55$, but the Rayleigh instability limits the maximum charge state for a given size $N$. The "large" HNDs considered here had $N/z = 7.5 \times 10^4$, hence $z_{max} = 4$. What is the contribution of fission fragments from doubly, triply, and quadruply charged droplets to the mass spectrum displayed in Fig. S2a?

In a doubly charged HND two singly charged Ne cluster ions will nucleate at the two $He_3^+$ charge centers. Using the procedure described in the main text (Section 4.2), with $N_i = 1.5 \times 10^5$, one would predict that the droplet shrinks to $N_f = 6.9 \times 10^4$ as it traverses the pickup cell. This value, however, is below the critical size $N_2 = 1.0 \times 10^5$, hence the droplet will undergo fission before reaching the exit of the pickup cell. Triply and quadruply charged HNDs will suffer the same fate.

Our aim is to model the growth of $z$ distinct $Ne_s^+$ ions in $He_N^{z+}$, the location of the fission reaction, the identity of its product ions, and their evolution during the remaining path. In previous experiments we have investigated highly charged HNDs that were mass-to-charge (hence size-to-charge) selected in a spherical electrostatic analyzer. The ions were post-ionized by another electron beam and the resulting $N/z$ distribution was measured in a second electrostatic analyzer. The $N/z$ values of fission fragment ions were all rational fractions. The observation implies that the ejected singly charged helium cluster ions are very small, containing fewer than ≈100 atoms, in agreement with conclusions drawn from more indirect experimental evidence.[2]

The model developed for the reactions in the pickup cell in Section 4.2 still applies to multiply charged HNDs except that the number $ds$ of Ne atoms added to each embedded $Ne_s^+$ after $dx$ collisions will now equal $dx/z$. Furthermore, $z$ changes to $z-1$ once the droplet size falls below the critical size $N_z$. One finds that an initially doubly or triply charged HND will experience one fission event in the cell; a quadruply charged HND will experience two successive fission events. Table S1 lists the charge state $z_b$, number of He atoms $N_b$ and size $s_b$ of the $Ne_s^+$ that are embedded in doped HNDs exiting the pickup cell, for each initial charge state $z_a$.

The fate of the one- or two-fold doped HNDs that enter the collision cell can be modeled as described in Section 4.3; the results are compiled in Table S1 for each initial charge state $z_a$. The droplets will shed all their He atoms and, along the way, fission into singly charged ions. The bare $Ne_s^+$ product ions will evaporate more than 50 % of their atoms before they reach the exit of the collision cell but their final sizes $s_c$ are somewhat larger than the final size computed for HNDs that are initially singly charged.

**Table S1**
Modeling growth and shrinkage of neon cluster ions. $N_a$ and $z_a$ are the size and charge state of HNDs that enter the pickup cell. Upon exiting the pickup cell they will contain $N_b$ He atoms and $z_b$ charges which reside on the embedded $Ne_{sb}^+$ cluster ions. $z_b < z_a$ implies the occurrence of fission. In the collision cell the HNDs will be stripped of the remaining He and, if $z_b = 2$, fission once more, resulting in bare cluster ions $Ne_{sc}^+$ at the exit of the collision cell. Listed in separate rows are the final sizes $s_c$ of the nearly bare neon cluster ions that are ejected upon fission from multiply charged HNDs.

| Entrance and exit of pickup cell | | | | | Exit of collision cell | |
|---|---|---|---|---|---|---|
| $z_a$ | $N_a$ | $z_b$ | $N_b$ | $s_b$ | $z_c$ | $s_c$ |
| 1 | 75000 | 1 | 29150 | 281 | 1 | 87 |
| 2 | 150000 | 1 | 73300 | 319 | 1 | 137 |
| | | | | | 1 | 10 |
| 3 | 225000 | 2 | 122000 | 273 | 1 | 119 |
| | | | | | 1 | 2, 63 |
| 4 | 300000 | 2 | 173200 | 260 | 1 | 123 |
| | | | | | 1 | 2, 38, 66 |

## S4 The Fate of Singly Charged Fission Fragments

Each fission event produces a pair of ions. In Section S3 we considered the fate of the *large* fission fragments that carry $z$-1 charges and almost all of the He atoms. In order to model the fate of their partners, nearly bare $Ne_s^+$ ions complexed with < $10^2$ He atoms, we need to track their size and the location where they are formed. This information is a byproduct of the modelling described in Section S3; results are compiled in Table S2. $L_1$ denotes the distance from the entrance to either the pickup cell or the collision cell, depending on where the ion is formed.

First we consider the fate of the bare $Ne_s^+$ ions that are produced in the pickup cell. The few He atoms that they contain will be neglected; a single collision would provide enough energy to completely evaporate them. The collision cross section of $Ne_s^+$ colliding with neon gas equals

$$\sigma = \pi R_{Ne}^2 (s^{1/3} + 1)^2 \tag{S2}$$

Each collision transfers an energy

$$E^* = 2k_B T + m v_d^2/2 = 74.0 \text{ meV} \tag{S3}$$

to the cluster ion; the colliding Ne atoms will not condense at the collisionally heated $Ne_s^+$. The relation between the change in cluster size $ds$ and the number of collisions $dx$ is

$$ds/dx = -(E^*/D_{Ne}) = -3.70 \tag{S4}$$

The remaining steps are similar to those leading to eq. 16; one finds

$$(s^{1/3} + 1)^{-2} ds = -n\pi R_{Ne}^2 \frac{v_r}{v_d} \frac{E^*}{D_{Ne}} dL \tag{S5}$$

with $n = 1.98 \times 10^{19}$ m$^{-3}$ ($P_{Ne} = 0.082$ Pa), $R_{Ne} = 0.1744$ nm, $v_r = 727$ m/s, and $v_d = 462$ m/s. Numerical integration of the right-hand-side over the remaining path from $L_1$ to $L = 4.45$ cm and the left-hand-side from the known initial size $s$ to the unknown final size $s_b$ at the exit of the pickup cell provides the value of $s_b$. Results are compiled in Table S2.

Eq. 20 still holds for the shrinkage of $Ne_s^+$ upon collisions with helium in the collision cell,

$$(s^{1/3} + R_r)^{-2} ds = -n\pi R_{Ne}^2 \frac{v_r}{v_d} \frac{E^*}{D_{Ne}} dL \tag{S6}$$

with $n = 2.69 \times 10^{19}$ m$^{-3}$ ($P_{He} = 0.112$ Pa), $R_r = R_{He}/R_{Ne}$, $v_d = 462$ m/s, $v_r = 1342$ m/s, and $E^*/D_{Ne} = 2.81$ Ne atoms. The right-hand-side is to be integrated over the length $L = 26$ cm of the collision cell if fission occurred in the pickup cell, or over the remaining length from $L_1$ to $L$ if it occurred in the collision cell.

The final sizes $s_c$ of ions that exit the collision cell cover a large range. The very first fission products of triply and quadruply charged HNDs contain few Ne atoms, and they will have to travel a long distance towards the exit of the collision cell. They will end up as tightly bound $Ne_2^+$ which will not dissociate further at 300 K. The other small fission fragments emerge with an average size ranging from 10 to 65.

**Table S2**
The fate of $Ne_s^+$ fragment ions that are ejected from a multiply charged doped HNDs. $He_{Na}^{za+}$ is the original ion that enters the pickup cell. The droplet will fission as its size decreases below the critical size upon collisions in the pickup or collision cell. The products are a HND that contains $N$ He atoms and $z$ distinct $Ne_s^+$ cluster ions, plus a nearly bare $Ne_s^+$. A total of $z_a - 1$ fission reactions will happen; they are listed in separate rows. They occur at a distance $L_1$ from the entrance of the cell that is specified in the following column. $s_b$ and $s_c$ specify the size of the bare neon cluster ions when they exit the pickup and collision cell, respectively.

| $z_a$ | $N_a$ | $z$ | $N$ | $s$ | $L_1$ (cm) | cell | $s_b$ | $s_c$ |
|---|---|---|---|---|---|---|---|---|
| 2 | 150000 | 1 | 100000 | 145 | 2.65 | pickup | 137 | 10 |
| 3 | 225000 | 2 | 184000 | 71 | 1.58 | pickup | 63 | 2 |
|  |  | 1 | 100000 | 273 | 0.84 | collision | - | 63 |
| 4 | 300000 | 3 | 283000 | 9 | 0.52 | pickup | 5 | 2 |
|  |  | 2 | 184000 | 225 | 4.01 | pickup | 223 | 38 |
|  |  | 1 | 100000 | 260 | 2.45 | collision | - | 66 |